# Aluminum oxide – n-Si field effect inversion layer solar cells with organic top contact

A.S. Erickson, N.K. Kedem, A.E. Haj-Yahia, D. Cahen

Dept. of Materials and Interfaces, Weizmann Institute of Science, Rehovot, 76100, Israel

We demonstrate a solar cell that uses fixed negative charges formed at the interface of n-Si with $Al_2O_3$ to generate strong inversion at the surface of n-Si by electrostatic repulsion. Built-in voltages of up to 755 mV are found at this interface. In order to to harness this large built-in voltage, we present a photovoltaic device where the photocurrent generated in this inversion layer is extracted via an inversion layer induced by a high work function transparent organic top contact, deposited on top of a passivating and dipole-inducing molecular monolayer. Results of the effect of the molecular monolayer on device performance yield open-circuit voltages of up to 550 mV for moderately doped Si, demonstrating the effectiveness of this contact structure in removing the Fermi level pinning that has hindered past efforts in developing this type of solar cell with n-type Si.

The standard process for producing commercial silicon solar cells is highly energy intensive, requiring about 1.5 GJ per panel to produce polycrystalline panels of area 0.65 $m^2$.[1] A major hindrance to reducing the cost of silicon solar cells is the high temperature dopant diffusion step in forming the emitter. In addition to the large amount of energy required to maintain a temperature of 900 C for multiple hours, in order to diffuse dopants from the surface into a Si wafer, such high temperature requirements preclude the use of inexpensive methods of forming the Si base using substrates such as glass or metal foils. Inversion layer solar cells circumvent this high temperature step by replacing the conventional doped p-n junction with a p-n junction induced in the top region of the substrate, using low temperature surface treatments alone.

An inversion layer can be formed by contacting an n- (p-) type semiconductor with a high (low) work function metal, forming a Schottky barrier that is high enough so that minority carriers become dominant in a narrow layer at the surface. This occurs at a built-in voltage of $V_{bi} > (E_G/2 - |E_C-E_F|)$, for an n-type semiconductor in contact with a high work function metal, where $E_G$ is the semiconductor band gap energy and $E_C$ is the conduction band edge energy. This effectively forms a p-n homojunction, while avoiding the energy-intensive processing steps, involved in doping. If the Schottky barrier is high enough that the carrier concentration in the inversion layer exceeds the dopant density, the junction is in strong inversion, and generally exhibits higher mobility due to improved screening of impurities. In practice, however, for Si and other covalent semiconductors, Fermi level pinning limits the actual barrier height of these types of devices, unless adequate surface passivation can be provided.[2]

Instead of using a metallic top contact to generate inversion in Si, fixed charges held at the interface of Si with a dielectric, such as $Al_2O_3$ or $SiO_x$, can generate inversion by electrostatic repulsion, if the majority carrier type of the Si substrate is of the same sign as the fixed charge at the interface, and if the density of fixed charges is high enough. This reduces the impact of Fermi level pinning by taking advantage of dielectrics that also passivate the Si surface. The most effective solar cells based on this concept were made in the SiO - p-Si system, where positive fixed charges at the Si-SiO interface generate inversion near the surface of the Si substrate. Devices made from thin SiO layers on p-Si, using thin-oxide tunnel junction contacts, reached 16.8% efficiency (active area), for Czochralski-grown substrates, implying 12.6% total device area efficiency, given the reported 25% shading.[3] The interface between $Al_2O_3$ and Si contains up to $10^{13}$ cm$^{-3}$ fixed negative charges, an order of magnitude higher than the positive charge density found at the SiO/Si interface,[4,5] which lead to field effect passivation in p-type and inversion in n-type Si.[5,6] The higher charge density at this interface should lead to stronger inversion than that found in the previously investigated SiO-Si system, and potentially, also to better device performance. However, making contact to such a device using an n-Si/oxide/metal junction, as was used in the p-Si/SiO system described above,[3] will not work as well, due to the low open circuit voltages typically found for metal-insulator-n-Si solar cells.[7,8] For this reason, n-Si inversion layer solar cells have not been extensively studied in the literature. Instead, metal-Si junctions, where the Si surface has been modified and passivated by a molecular monolayer to suppress Fermi level pinning, show more promise in generating the high $V_{oc}$ needed for these devices, while maintaining the desired low processing temperatures.[9]

In this letter, we demonstrate inversion layer solar cells based on the negatively charged interface between $Al_2O_3$ and n-Si. Figure 1(a) shows a proposed band diagram of this interface (adapted from ref. 6). We show that the negative fixed charge density leads to a built-in voltage ($V_{bi}$) of up to 755 mV, more than sufficient to reach strong inversion for a substrate donor concentration, $N_D$, less than $5\times10^{16}$ cm$^{-3}$. As a thickness of at least 10-15 nm $Al_2O_3$ is required to achieve a high interface charge density,[5] contact points to extract carriers must be etched through the $Al_2O_3$, which naturally removes inversion at the point of contact. A selective contact must then be formed, in order to avoid shorting to the n-type substrate. This can be done by diffusing p-type dopants at the contacts, which would create a conventional doped p-n junction at the contact. However, the same high temperatures we seek to avoid would be required for this step. Instead, we use a high work function top contact to generate inversion with room temperature processing. To remove Fermi level pinning, we first deposit a molecular monolayer (MM) of a strongly electron-donating molecule, ethylene toluene, to provide electronic passivation, as well as to introduce a dipole that can modify the electron affinity of the underlying Si, to increase the Schottky barrier height.[9,10] We then apply the organic conductor poly(ethylenedioxy thiophene):poly(styrenesulfonate) (PEDOT:PSS) as a high work function top contact, using a soft-deposition method that preserves the MM. We show that this contact structure leads to strong inversion for a substrate donor concentration, $N_D < 10^{15}$ cm$^{-3}$, though $V_{bi}$ in this configuration is 50-100 mV lower than that under the $Al_2O_3$ interface, and demonstrate the impact of the dipole-inducing MM at this interface on device performance.

The device structure is diagrammed in figure 1(b). In this configuration, under illumination, electron-hole pairs are primarily generated underneath the strong inversion layer created by the $Al_2O_3$ – n-Si interface and separated by the large $V_{bi,}$ present in this region. The hole is then transported through the inversion layer to the MM-passivated PEDOT:PSS contact, while the electron reaches the back contact via the Si substrate. Using this technique to create a p-n homojunction in n-Si, while maintaining processing temperature below 450 °C, we reach a best total area power conversion efficiency of 8.4%, with an average over 6 devices of 7.9(4)%. The lowering of required temperatures allows cheaper substrates to be used, making this method compatible with recent developments in forming inexpensive thin-film Si on glass substrates,[11] in addition to lowering the total amount of energy required to form the emitter. In this letter, single crystal Si is used as a model system, to better characterize the relevant interfaces and the resulting device performance trends.

Aluminum oxide was deposited by plasma-enhanced atomic layer deposition (PEALD) in a Cambridge NanoTech Fiji F200 system at a substrate temperature of 250 °C on 400-500 μm thick, single side polished, (111) oriented, P-doped n-type, Czochralski-grown single crystal Si wafers that had been thoroughly degreased and etched to leave a clean, smooth, and electronically well-passivated Si-H surface. The sample was then annealed for 30 min at 425 °C, to reduce the interface trap density.[12] Film thickness was measured by spectroscopic ellipsometry, and confirmed by TEM, where a 3 nm layer of $SiO_x$ was found at the interface of Si with $Al_2O_3$, typical for these deposition conditions.[5] Contact fingers of width 15 μm and spacing 50-150 μm, with a total device area 0.5 cm x 0.5 cm, were patterned by photolithography onto the $Al_2O_3$ films, and the exposed $Al_2O_3$ was then etched away. A MM of ethylene toluene was deposited by a method described elsewhere,[10] using ethynyl toluene as a molecular precursor, to passivate the exposed Si surface before depositing the top contact. The processing required for MM deposition resulted in a reduction of the $Al_2O_3$ thickness from 40 nm to 25 nm, due to etching steps used to prepare at the contacts the Si-H surface that is required to achieve a high quality MM. The presence of $Al_2O_3$ in these devices required a reduction in etching time from the optimal 15 min[10] to 2 min, to preserve the quality of the $Al_2O_3$ layer, a change that was found to affect device performance, as described below. MM characterization is described in reference 10. A 200 nm thick film of PEDOT:PSS (Pedot P, H.C. Starck), doped with 5% ethylene glycol and 0.4% zonyl surfactant was deposited by spin-coating onto a sacrificial glass substrate and annealed for 30 min at 150 °C in a nitrogen environment. This film was then deposited onto the device by the lift-off-float-on (LOFO) method,[13,14] preventing exposure of the MM to the acidic, aqueous PEDOT:PSS solution. Devices made by similar methods but with direct deposition of spin-coated PEDOT:PSS were not as efficient. A 200 nm thick Ag collection grid, contributing 15% shading, was deposited by e-beam evaporation, and In-Ga eutectic was used as the back contact. Photovoltaic performance was evaluated using two-probe I-V techniques, with a 100 mW/cm² W-halogen lamp, calibrated using a commercial silicon diode. Two of the best devices were also measured under a solar simulator to verify the efficiency. A 3% increase in the short circuit current was found, relative to the halogen lamp, and is likely due to improved transparency of PEDOT:PSS in the UV, where the actual solar spectrum is stronger that of the halogen lamp.

The inversion layer formed by the $Al_2O_3$ - n-type Si interface was characterized on Si with a 30 nm thick $Al_2O_3$ layer, by capacitance-voltage (C-V) profiling and by Kelvin probe surface photovoltage (SPV) measurements. The fixed charge density at the interface, $Q_f$, was determined from the flat-band voltage, $V_{fb}$, measured from the C-V profile, taken at an applied frequency of 1 MHz, by the relation $V_{FB} = \varphi_{MS} - \frac{Q_f}{C_{ox}}$, where $\varphi_{MS}$ is the difference in Fermi energy between the metal contact and semiconductor substrate, and $C_{ox}$ is the oxide capacitance, measured by C-V.[15] The interface trap density, $D_{it}$, referring to charge states in the $Al_2O_3$ layer that interact electrically with mobile carriers in the Si substrate, was determined for $N_D$ = 5 x $10^{14}$ cm$^{-3}$ by comparing data taken at 100 Hz and at 1 MHz.[16] This measurement was not done for samples with $N_D$ = 5x$10^{13}$ cm$^{-3}$ due to spurious effects at low frequency, thought to arise from poor contact to the In-Ga eutectic at low substrate $N_D$. The built-in voltage, $V_{bi}$, was measured by SPV. The carrier concentration in the inversion layer, p(IL), was then determined from the measured $V_{bi}$ and the offset between $E_F$ and the conduction band edge, $E_C$, in the bulk Si. For the case of $N_D$ = 5 x $10^{14}$, the built-in voltage was measured by C-V to be 755 mV, 100 mV larger than that found by SPV. Due to inherent limitations in determining the photosaturation limit in SPV measurements,[17] the value of $V_{bi}$ derived from SPV, presents a lower limit, and the value from C-V is considered more accurate. However, as C-V at low frequency was not possible for the sample with lower doping, $V_{bi}$ measured by SPV was used to compare p(IL). Additionally, $V_{bi}$ at the PEDOT:PSS – MM – n-Si contact was measured by SPV in two test devices with no $Al_2O_3$, with $N_D = 5 \times 10^{14}$, and was found to be 50-100 mV lower than at the $Al_2O_3$ – n-type Si interface, when also measured by SPV.

The fixed charge density, $Q_f$, measured by C-V, reached $3.2(5) \times 10^{12}$ cm$^{-2}$, with the interface trap density, $D_{it}$, as low as $10^{11}$ cm$^{-2}$, near the values expected for high quality $Al_2O_3$ passivation layers, reported in the literature.[5] The resulting built-in voltage, $V_{bi}$, measured by SPV and, for the case of $N_D = 5 \times 10^{14}$ cm$^{-3}$, by C-V methods, is shown in figure 2 (right axis). The carrier concentration near the surface, derived from $V_{bi}$, indicates inversion in all cases studied, and strong inversion for $N_D < 5 \times 10^{16}$ cm$^{-3}$ (fig. 2, left axis). Additionally, low frequency C(V) behavior was seen as high as 100 kHz, which lends further proof to our claim of strong inversion.[15] Details are presented in supplemental information, available online.[18]

Photovoltaic performance results are shown in figure 3, where devices of different $N_D$ are compared, and the results are summarized in Table I. For higher doping substrates, with $N_D = 5 \times 10^{15}$ cm$^{-3}$, the built-in voltage measured by SPV indicates that the devices are near the strong inversion limit. These devices show poor performance at 150 μm and 100 μm contact spacing, yielding significant photocurrent only at the narrowest spacing, 50 μm, where the fill factor remained low. The s-shape near $V_{oc}$ is characteristic of a reverse diode, which in this device may lie at the junction between the stronger inversion layer at the $Al_2O_3$-Si interface, and the weaker one induced by the PEDOT:PSS. Holes traveling from the higher to lower doping when moving from the strong IL under the $Al_2O_3$ to the contact will experience a transport barrier. When $N_D$ is reduced to $5 \times 10^{14}$ cm$^{-3}$, where strong inversion is found over the entire device area, significant improvements in all performance parameters are observed. In this case, devices with 50 μm or 100 μm contact spacing showed equally good performance (figure 3, red solid line), with average efficiency reaching 7.9(4)% and a maximum of 8.4%. This suggests that the mobility in this inversion layer, which is expected to increase with increasing inversion strength due to improved screening of donor ions, is sufficient to extract carriers over the larger distances. When $N_D$ is further reduced to $5 \times 10^{13}$ cm$^{-3}$, an improvement in short circuit current shows the effect of the strongest inversion layer, and devices with contact spacing up to 150 μm showed good performance. However, the reduced fill factor, likely due to the increased series resistance of the substrate, reduces overall efficiency.

Several types of control devices were made using wafers with $N_D = 5 \times 10^{14}$ cm$^{-3}$, to assess the relative impact of different factors in device fabrication. Figure 3 (green dash-dot line) shows the performance of a device with the same structure as the devices described above, without the passivating molecular monolayer. Instead, the PEDOT:PSS was deposited by the same method on a Si-H terminated contact. In this case, a reverse diode was present in all samples measured. This is consistent with the proposed effect of the molecular dipole, which should strengthen the inversion layer at the PEDOT:PSS contact, reducing the potential barrier required for holes to pass to the contact.

Two types of devices without Al$_2$O$_3$ were also studied, comprising a solar cell based solely on the n-type Si – MM - PEDOT:PSS (deposited by the LOFO method) junction. In one case, the restricted 2 min NH$_4$F etch time required for the Al$_2$O$_3$ devices was used, to measure the effect of these limited etching conditions on device performance. In this case, only a 5.9(8)% average efficiency was found. If, however, optimal 15 min NH$_4$F etch time is used in preparing the surface for the MM passivation,[10] a device with only n-type Si – MM - PEDOT:PSS structure reached an average efficiency of 7.4 ± 1.0%. The large error bar reflects the relatively poor reproducibility found for these devices, compared to the Al$_2$O$_3$-based devices. Additionally, spin-coating PEDOT:PSS directly onto Si-H, using similar conditions to those provided in literature,[19] led to an efficiency of only 7%. In all of the devices studied, the use of the In-Ga eutectic back contact is likely to impact efficiency through increased series resistance, compared to diffused back contacts.

External quantum efficiency and reflectivity were also compared for a device with $Al_2O_3$, and for a PEDOT:PSS only device, prepared using the optimized sample preparation conditions, and is presented in supplementary material, available online.[20] The enhancement in photocurrent for devices with $Al_2O_3$ was found to derive primarily from long wavelength (above 800 nm) and short wavelength (below 500 nm) regions, likely resulting from reduced reflectivity, and from improved collection in the bulk due to a wider depletion width arising from the stronger inversion layer, respectively.

In summary, we demonstrated that the negatively-charged interface of $Al_2O_3$ with n-type Si produces large built-in voltages of up to 755 mV, 50-100 mV higher than that induced by PEDOT:PSS, leading to strong inversion for substrate doping concentrations less than $10^{15}$ cm$^{-3}$. This built-in voltage was used as the primary charge separation potential in a solar cell, using contacts based on the interface of n-Si, passivated with a dipole-inducing molecular monolayer, with PEDOT:PSS as a high work function conductor. By comparing to devices with Si-H passivated contacts, we demonstrated the effect of the molecular dipole in enhancing inversion at the contacting interface. An average device efficiency of 7.9(4)% was reached, exceeding that of PEDOT:PSS - only devices, consistent with the larger built-in voltage found. $Al_2O_3$-based devices also showed vastly improved reproducibility over PEDOT:PSS – n-Si solar cells, suggesting that the inversion layer generated by the fixed charges is more robust to process variation than that induced by the organic conductor. Comparing optimized to actual MM deposition conditions suggests that improvements in contact processing can yield further device improvements, and indications from C-V measurements at low frequency indicate that an improved back contact may also be helpful.

We thank the Helmsley trust, the Wolfson family trust, and the Grand Center for Sensors and Security for partial support. We thank K.L. Narasimhan (TIFR, India) for support with C-V measurement and analysis, Ellen Moons (Karlstad Univ., Sweden) and Rotem Har-Lavan (Weizmann Institute of Science) for instructive conversations and the reviewer for suggesting the reflectivity experiment. ASE acknowledges a Weizmann Institute dean's postdoctoral fellowship. DC holds the Schaefer Chair in Energy Research.

*Express* **5,** 032301 (2012); T. –G. Chen, B.-Y. Huang, E.-C. Chen, P. Yu, and H. –F. Meng, *Applied Physics Letters* **101**, 033301 (2012)).

20. See supplementary material at [URL will be inserted by AIP] for detailed description of external quantum efficiency measurements.

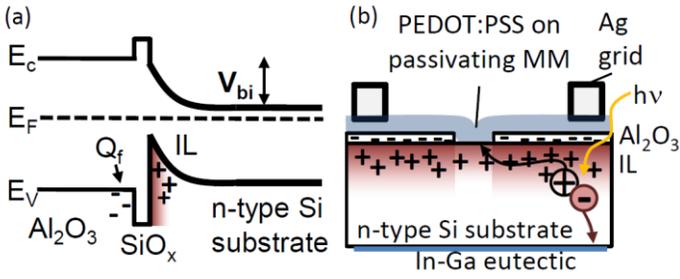

Figure 1 (color online). (a) Proposed band structure of $Al_2O_3$ – n-type Si interface (adapted from reference 5), showing the origin of the built-in voltage, $V_{bi}$. The interfacial SiOx layer arises during $Al_2O_3$ deposition. (b) Scheme of final device structure, showing proposed transport path for photogenerated carriers.

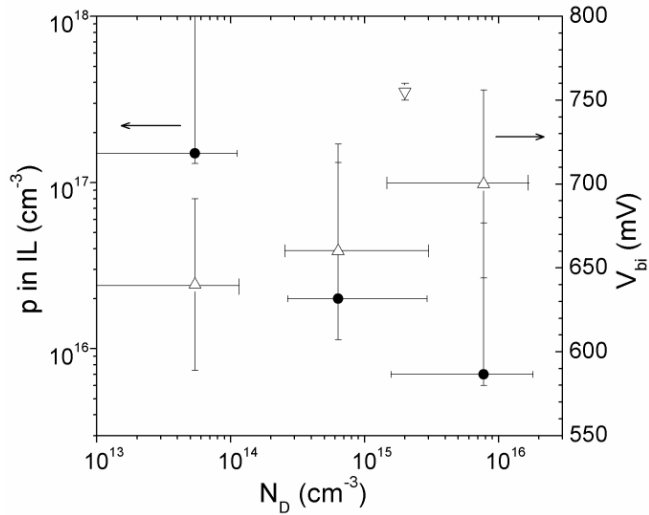

Figure 2. Built-in voltage, $V_{bi}$ (open triangles, right axis), measured by SPV (up triangles), and by C-V (down triangle), and hole concentration in the inversion layer, p (filled circles, left axis), calculated from $V_{bi}$. Vertical error bars represent instrumental limits, error bars in $N_D$ stem from the doping range given for the source wafers, except for the case of the sample measured by C-V (down triangle), where $N_D$ was extracted from the C-V measurement and the error is smaller than the symbol.

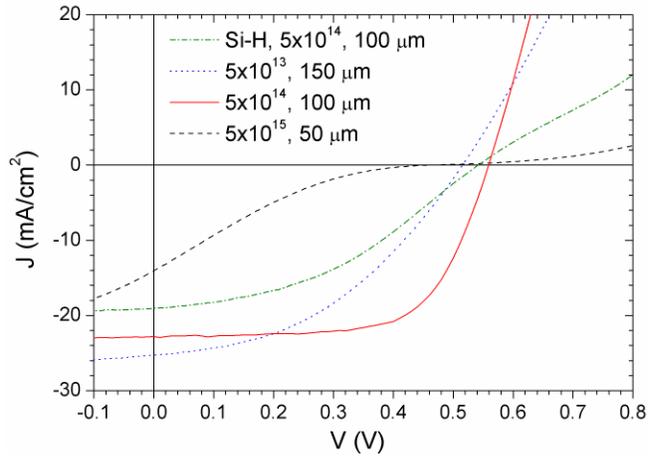

Figure 3 (color online). J(V) data, taken under 100 mW/cm$^2$ light intensity for devices with different substrate dopant concentrations and contact spacing, as noted. The sample labeled "Si-H" refers to the device with $N_D = 5 \times 10^{14}$ cm$^{-3}$ and 100 μm contact spacing, with no passivating MM.

Table I. Summary of photovoltaic device performance for $Al_2O_3$-based devices with (lines 1-3) and without (line 4) passivating MM, PEDOT:PSS-only control device with the same process conditions as $Al_2O_3$ devices (line 5) and optimized PEDOT:PSS-only device (line 6) both with $N_D = 5 \times 10^{14}$ $cm^{-3}$. Results shown are averaged over 3-6 best devices of each type, with the standard deviation indicated.

| Sample | $J_{sc}$ (mA/cm$^2$) | $V_{oc}$ (V) | FF (%) | PCE (%) |
|---|---|---|---|---|
| $5 \times 10^{13}$ | 23(2) | 0.511(3) | 0.52(6) | 6.1(5) |
| $5 \times 10^{14}$ | 23(1) | 0.553(6) | 0.63(6) | 7.9(4) |
| $5 \times 10^{15}$ | 14(2) | 0.48(2) | 0.165(7) | 1.1(2) |
| $5 \times 10^{14}$, Si-H | 18(1) | 0.535(7) | 0.47(9) | 4.6(6) |
| PEDOT control | 20(3) | 0.503(6) | 0.60(7) | 5.9(8) |
| PEDOT optimized | 21(1) | 0.51(2) | 0.67(5) | 7.4(1.0) |

**Aluminum oxide – n-Si field effect inversion layer solar cells with organic top contact**

A.S. Erickson,[1] N.K. Kedem,[1] A.E. Haj-Yahia,[1] D. Cahen[1]

[1]Dept. of Materials and Interfaces, Weizmann Institute of Science, Rehovot, 76100, Israel

For the case of a typical metal-oxide-semiconductor (MOS) capacitor, low frequency capacitance-voltage (C-V) curves are generally characterized by reverse bias capacitance that saturates at values similar to the oxide capacitance, due to the response of minority carriers in the inversion layer.[1] The frequency range, where this type of behavior is normally seen, depends on specific parameters of the system, but typical values for an inverted Si surface are up to tens of Hz. At higher frequency, the reverse bias capacitance typically saturates at low values due to the relatively long response time of the minority carriers. The highly inverted Si-$Al_2O_3$ interface, found in this work, gives rise to a saturation of the high frequency capacitance at reverse bias, similar to the behavior found only at low frequency for samples with a weaker inversion layer. This type of behavior indicates the existence of a uniform inversion layer at zero bias, which extends far beyond the gate electrode.[1] This strong inversion layer, which acts as a good conductor, allows AC current to pass through an area much larger than the gate. The capacitance of the system is then governed by the capacitance of the depletion layer in that larger area, until it saturates at the oxide capacitance. Stronger inversion implies lower resistance in the inversion layer, due to the higher carrier density, which allows this typically low-frequency behavior to persist at higher frequencies. In figure 1, a C-V plot of a typical Au - $Al_2O_3$ - n-Si MOS device at various frequencies is presented. The capacitance at reverse bias is similar to the oxide capacitance at relatively high frequencies of up to 100 kHz, and the impact of this highly conductive inversion layer is noticeable even at 1 MHz, in the slight upturn at reverse bias. Figure 2 shows the reverse bias capacitance at different frequencies. The reverse bias capacitance saturates at the oxide capacitance as high as 10 KHz. The fact that this type of behavior persists at such high frequency further supports our claim of strong inversion at the Si-$Al_2O_3$ interface.

External quantum efficiency (EQE) was measured using a 550 W tungsten-halogen lamp with a monochromator and a mechanical chopper to produce the incident light spectrum, which was calibrated to a silicon diode with known EQE. Current was measured using a lock-in amplifier locked-in to the frequency of the chopped light. For data shown here, the chopping frequency was 10 Hz, and no bias light or voltage were applied. Varying frequency and bias light intensity was found not to affect the results. Complementary reflectivity measurements were taken using a Jasco V-570 spectrophotometer with an integrating sphere. Samples of Si – $Al_2O_3$ – PEDOT:PSS, with varying $Al_2O_3$ thickness and fixed 0.4 x 0.6 cm size, were loaded at an angle to minimize light reflected directly back to the source.

External quantum efficiency results for samples of roughly average efficiency are shown in figure 3(a), comparing a standard-process $Al_2O_3$ device to a device with PEDOT:PSS only (using optimized processing conditions, as described in the main text) to understand the source of improved photocurrent for $Al_2O_3$-based devices. In this case, both samples had a similar $J_{sc}$ of 21.1 mA/cm$^2$ for the $Al_2O_3$ device, and 21.5 mA/cm$^2$ for the PEDOT-only device. As is more clearly seen when the spectra are normalized to the maximum value (Fig 3(b)), the $Al_2O_3$ provides enhancements to photocurrent collection primarily in the UV and IR regions.

Results of reflectivity measurements are shown in figure 4. A reduction in reflectivity is seen below 500 nm for samples of $Al_2O_3$ thickness near the 25 nm used in this study, corresponding to the region of enhanced quantum efficiency in the UV. Comparing to data for a much thicker $Al_2O_3$ layer of 45 nm shows that this arises from interference fringes. This raises the possibility of using a much thicker layer of $Al_2O_3$, perhaps by combining ALD with a higher throughput method, to take advantage of the antireflective properties. The $Al_2O_3$ layer was not found to influence the long wavelength reflectivity.

The UV enhancement of the quantum efficiency likely results from reduced reflectivity in the presence of $Al_2O_3$, as demonstrated above. The improved IR response most likely arises from the wider depletion width expected for samples with a stronger inversion layer, which would improve collection of carriers generated in the bulk of the wafer. A higher efficiency $Al_2O_3$ device of PCE = 8.4% and Jsc = 23.0 mA/cm$^2$ was also measured and was found to show an even stronger IR and UV enhancement, as well as an overall higher EQE, but those data are not shown here, so as to emphasize that the differences are present also for average efficiency devices. The relatively low performance of the PEDOT:PSS – only device in the IR is typical for PEDOT:PSS – n-Si solar cells reported in the literature.[2] We note that the significantly higher $V_{oc}$ values found for $Al_2O_3$ devices cannot be explained by the increase in EQE, and much more likely arise from the larger built-in voltage found in the inversion layer, induced by the high charge-density interface.

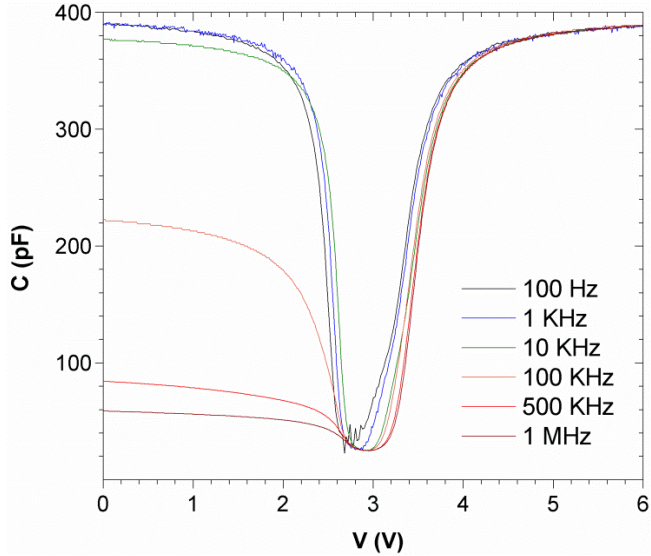

Figure 1. Capacitance as a function of gate bias voltage, for different applied frequencies. All plots are corrected to exclude the effect of substrate and back contact resistance.

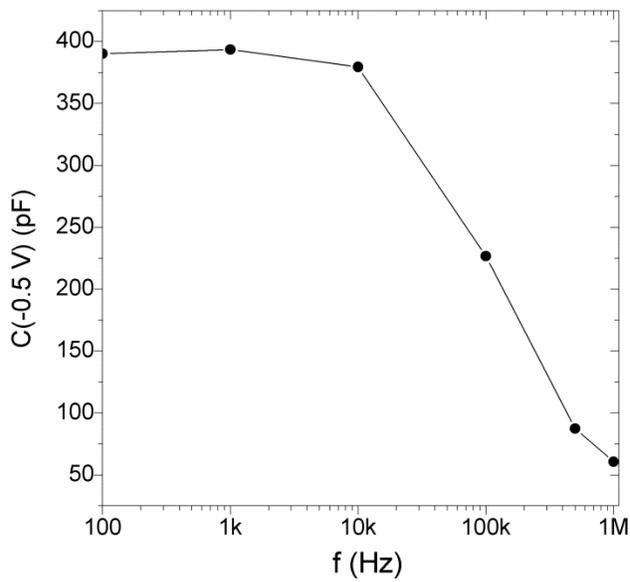

Figure 2. Capacitance of the MOS devices at -0.5 V bias vs. measurement frequency, showing the saturation of capacitance at frequencies up to 10 KHz.

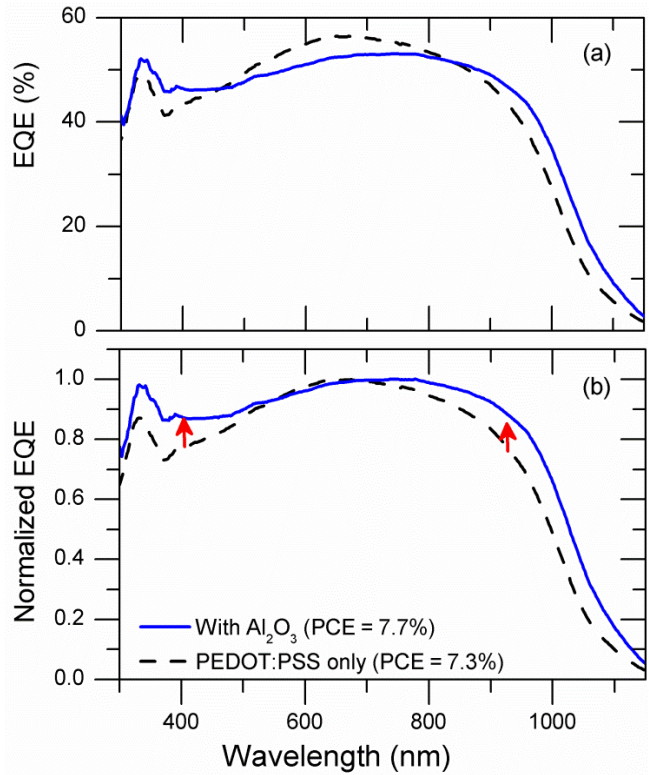

Figure 3. (a) External quantum efficiency, and (b) EQE normalized to the maximum value, for a standard device with $Al_2O_3$ and a control device with only n-Si – PEDOT:PSS, which was deposited using the optimized processing conditions. Power conversion efficiency (PCE) for these particular devices is also indicated.

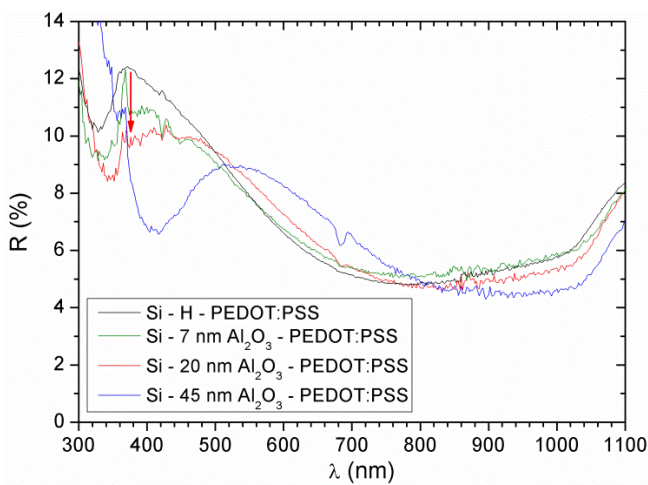

Figure 4. Reflectivity measurements for samples of n-Si with varying thickness $Al_2O_3$ layers and 200 nm PEDOT:PSS, and for n-Si – H without $Al_2O_3$, but with a 200 nm PEDOT:PSS, as a reference sample.